%% file: main.tex
\documentclass[10pt,twocolumn,letterpaper]{article}
\usepackage{multirow}
\usepackage{footmisc}
\usepackage[pagenumbers]{iccv}
\definecolor{iccvblue}{rgb}{0.21,0.49,0.74}
\usepackage[pagebackref,breaklinks,colorlinks,allcolors=iccvblue]{hyperref}

\newcommand{\OURS}{VertexRegen}

\newcommand{\customfootnotetext}[2]{%
  \begingroup
  \renewcommand{\thefootnote}{#1}%
  \footnotetext{#2}%
  \endgroup
}

\title{\OURS: Mesh Generation with Continuous Level of Detail}

\author{
Xiang Zhang$^1$\textsuperscript{*}
\quad Yawar Siddiqui$^2$
\quad Armen Avetisyan$^2$
\quad Chris Xie$^2$\\ 
\quad Jakob Engel$^2$
\quad Henry Howard-Jenkins$^2$\\
$^1$UC San Diego \quad $^2$Meta Reality Labs Research
}
\begin{document}

\input{figures/teaser}
\input{sec/0_abstract}
\customfootnotetext{*}{Work conducted while the author was an intern at Meta.}
\customfootnotetext{}{Project page: \url{https://vertexregen.github.io}}
\input{sec/1_intro}
\input{sec/2_related_work}
\input{sec/3_method}
\input{sec/4_experiments}
\input{sec/5_conclusion}

{
    \small
    \bibliographystyle{ieeenat_fullname}
    \bibliography{main}
}

\end{document}

%% file: figures/teaser.tex
\twocolumn[{%
    \renewcommand\twocolumn[1][]{#1}%
    \maketitle
    \begin{center}
    \vspace{-7mm}    
    \includegraphics[width=\linewidth]{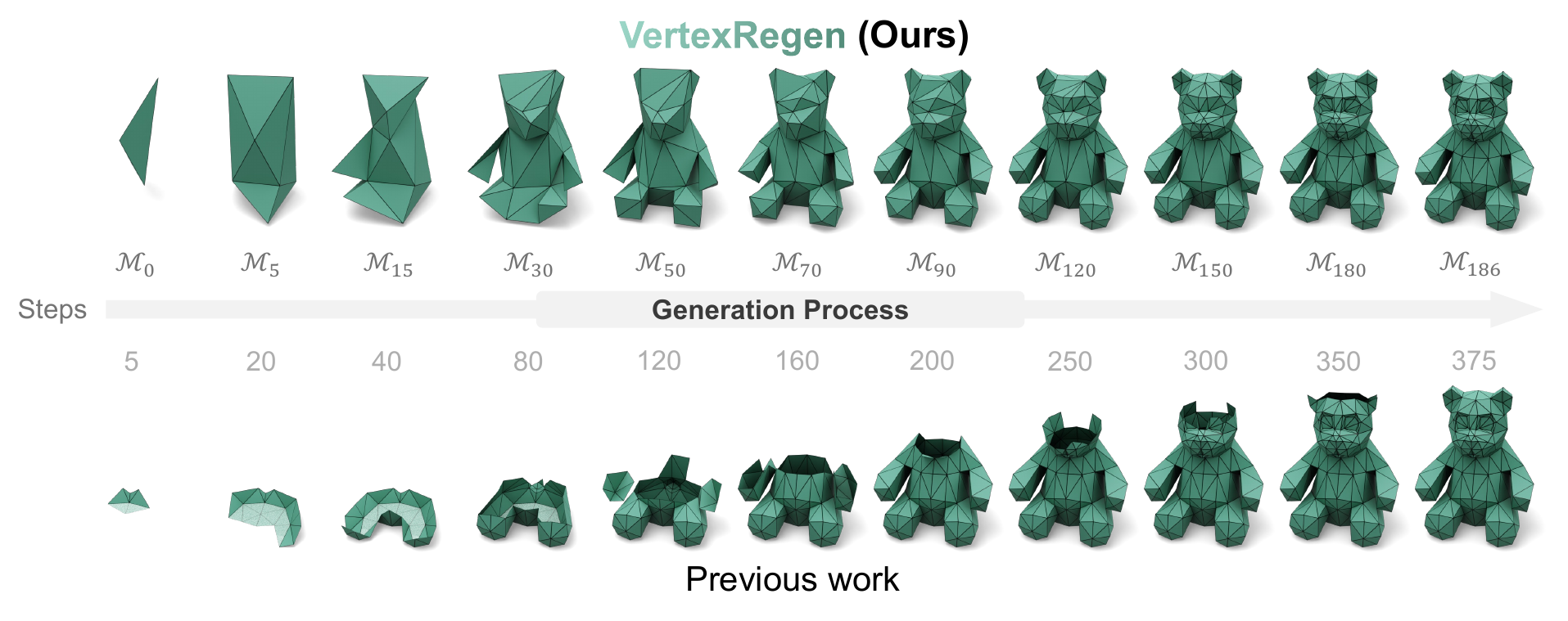}
    \captionof{figure}{Mesh generation process of \textbf{\OURS{}} (\textit{coarse-to-fine} process) \vs previous work (\textit{partial-to-complete} process). Prior work generates meshes face by face, with the step count corresponding to the face count in the figure. In contrast, \OURS{} produces meshes with a continuously increasing level of detail, where each step yields a valid mesh $\mathcal{M}_k$.}
    \label{fig:teaser}
    \end{center}
}]

%% file: sec/0_abstract.tex
\begin{abstract}
We introduce \OURS{}, a novel mesh generation framework that enables generation at a continuous level of detail. Existing autoregressive methods generate meshes in a partial-to-complete manner and thus intermediate steps of generation represent incomplete structures. \OURS{} takes inspiration from progressive meshes and reformulates the process as the reversal of edge collapse, \ie vertex split, learned through a generative model. Experimental results demonstrate that \OURS{} produces meshes of comparable quality to state-of-the-art methods while uniquely offering \textit{anytime} generation with the flexibility to halt at any step to yield valid meshes with varying levels of detail. 
\end{abstract}

%% file: sec/1_intro.tex
\section{Introduction}

Meshes are essential for 3D asset representation and are widely used in industries such as film, design, and gaming due to their compatibility with most 3D software and hardware. 
However, manually creating high-quality meshes is costly and time-consuming, prompting extensive research into automated 3D content creation. 
Largely, these approaches have used alternative representations such as neural fields~\cite{gao2022get3d, park2019deepsdf, mescheder2019occupancy, siddiqui2024assetgen}, voxels~\cite{wu2016learning, schwarz2022voxgraf} or point clouds~\cite{zhou20213d, zeng2022lion} which are later post-processed to meshes instead of direct mesh modeling.
Unfortunately, these post-processed meshes often exhibit poor topology, over-tessellation, and artifacts, lacking the quality of artist-crafted meshes.

Recently, there has been a surge of approaches that directly generate meshes, using autoregressive models to represent meshes as sequences of triangles~\cite{siddiqui2024meshgpt,chen2024meshxl, chen2024meshanything,chen2024meshanythingv2,hao2024meshtron,tang2024edgerunner}. 
These methods capture the high-fidelity and aesthetic qualities of artist-created works without the need for post-generation conversion.
Although significant progress has been made in improving tokenization schemes~\cite{siddiqui2024meshgpt,chen2024meshxl,chen2024meshanything} and network architectures~\cite{hao2024meshtron}, these approaches do not fundamentally alter a key characteristic of generation: namely, to produce a valid mesh, the full sequence must be generated to completion.
Consequently, these methods offer no control over the level of detail during generation; early stopping results in a mesh with missing faces. 
Simple extensions, such as face-count conditioning tokens in EdgeRunner~\cite{tang2024edgerunner}, have been proposed to introduce some coarse control over the detail by pre-specifying a target face count. 
However, each generation still must be completed in its entirety to yield a complete mesh, and thus a single generation still offers only a single level of detail.

To allow for a continuous level of detail generation, we take inspiration from Hoppe's progressive mesh formulation~\cite{hoppe1996progressive}. 
Progressive meshes use two reversible operations to transition between levels of detail. 
The edge collapse operation simplifies a mesh into a coarser one by reducing one edge at a time. The vertex split operation reverses this simplification to add more detail by using information stored during edge collapse. 
By starting with a coarse mesh and keeping edge collapse records, this approach creates an efficient, lossless representation of the original mesh, allowing for continuous resolution adjustments by applying any number of vertex split operations.

In this paper, we build on this progressive mesh representation by learning the vertex split, \ie reversing the edge collapse operation, as a generative problem. 
This allows the resulting mesh generation to inherit the properties of progressive meshes and provides a solution for \textit{anytime} mesh generation, where the process can be stopped early to yield a coarser mesh rather than an incomplete one. By properly serializing the vertex split sequence, the entire generation process can be modeled using a Transformer trained with a next-token prediction objective, a widely adopted paradigm in mesh generation. 
We evaluate \OURS{} on the task of unconditional mesh generation, demonstrating superior results both qualitatively and quantitatively. We further provide examples of shape-conditioned generation. These experimental results illustrate the ability of our method to generate compelling meshes at a continuous resolution.

Our contributions can be summarized as follows:

\begin{itemize}
    \item Inspired by Hoppe's progressive meshes~\cite{hoppe1996progressive}, we reframe mesh generation as the reversal of edge collapse operations, \ie generating vertex splits.
    \item We formulate a token-efficient parameterization of a progressive mesh, through a half-edge data structure.
    \item We propose \OURS{} for continuous level of detail mesh generation. \OURS{} creates meshes in a coarse-to-fine fashion, rather than partial-to-complete, uniquely providing a solution for \textit{anytime} generation. 
\end{itemize}

%% file: sec/2_related_work.tex
\section{Related Work}

\paragraph{3D Mesh Generation.} Recent advancements in 3D shape generation have explored various representations, including point-clouds~\cite{zhou20213d,nichol2022point,vahdat2022lion,xu2024bayesian}, signed distance functions (SDFs)~\cite{li2023diffusion,shim2023diffusion,cheng2023sdfusion,liu2024meshformer,long2024wonder3d,xu2024instantmesh,zhao2025depr}, neural radiance fields (NeRFs)~\cite{jun2023shap,zhang20233dshape2vecset,hong2024lrm,wei2024meshlrm}, and Gaussian splatting~\cite{tang2024dreamgaussian,guedon2024sugar}. These implicit representations require iso-surface extraction techniques~\cite{lorensen1998marching,shen2021deep,wei2023neumanifold} to output meshes, often resulting in over-tessellated and excessively smooth outputs, which pose challenges for downstream applications such as geometric processing and manipulation.

In contrast, direct mesh generation inherently produces structured, well-defined geometry without the need for post-processing or surface extraction, making it an increasingly prominent approach in recent years. Early methods tackle this task by generating meshes from surface patches~\cite{groueix2018papier}, deforming ellipsoids~\cite{wang2018pixel2mesh}, predicting mesh graphs~\cite{dai2019scan2mesh}, or employing binary space partitioning~\cite{chen2020bsp}. More recent techniques leverage generative models, particularly diffusion models and sequence-based approaches. PolyDiff~\cite{alliegro2023polydiff} applies discrete diffusion, while PolyGen~\cite{nash2020polygen} autoregressively predicts vertices and faces using two separate networks. The sequence modeling paradigm has been further refined by representing the entire mesh as a single sequence. MeshGPT~\cite{siddiqui2024meshgpt} introduces a tokenization scheme based on a vector quantized variational auto-encoder (VQ-VAE), while MeshXL~\cite{chen2024meshxl} directly models discretized triangle soup sequences without compression. MeshAnything~\cite{chen2024meshanything} extends MeshGPT by incorporating a point-cloud encoder for shape-conditioned generation.  PivotMesh~\cite{weng2024pivotmesh} introduces a hierarchical approach, generating pivot vertices before producing the full mesh. Further research has focused on optimizing tokenization and sequence modeling. Adjacency-aware compression techniques~\cite{chen2024meshanythingv2,weng2024scaling,tang2024edgerunner} improve tokenization efficiency and enable higher face counts within fixed context windows, while Meshtron~\cite{hao2024meshtron} leverages Hourglass Transformers and sliding window attention to scale MeshXL sequences more effectively.

Despite these advancements, most existing approaches follow a partial-to-complete paradigm, where mesh regions are constructed sequentially at a uniform level of detail. Our work instead adopts a coarse-to-fine approach, continuously increasing the level of detail as new tokens are generated, achieving better control over geometric complexity.

\paragraph{Level of Detail Representations.} Level of Detail~\cite{luebke2002level} (LOD) is a widely used technique in computer graphics to optimize rendering performance by reducing the complexity of 3D models based on their size, distance from the camera, or importance in a scene. Different LOD strategies depend on the underlying representation of 3D shapes. For meshes, progressive meshes~\cite{hoppe1996progressive} generate a mesh sequence starting from a coarse base model, gradually refining it through a series of transformations that incrementally add detail. Traditional mesh simplification methods~\cite{garland1997surface,lindstrom1998fast} can be adopted to construct such sequences by iteratively reducing polygon count while preserving geometric fidelity. Progressive simplicial complexes~\cite{popovic1997progressive} extend progressive meshes to handle arbitrary meshes, including non-manifold and non-orientable surfaces.

Recent studies have incorporated neural networks for LOD representations. Neural progressive meshes~\cite{chen2023neural} propose a learned approach supporting LOD with a subdivision-based encoder-decoder. Similar representations have been explored for reconstruction approaches with signed distance functions (SDFs)~\cite{tang2018real,tang2020deep,takikawa2021neural} and neural fields~\cite{mujkanovic2024neural}. However, these approaches are not generative; they typically encode and decode existing meshes~\cite{chen2023neural} or optimize the LOD representation per shape~\cite{takikawa2021neural}. In contrast, our approach is purely generative. VertexRegen learns to create meshes from scratch, progressively adding detail through autoregressive sequence generation. This allows us to generate new, high-fidelity meshes without relying on pre-existing structures, distinguishing our method from both traditional and neural LOD techniques.

%% file: sec/3_method.tex
\section{Method}

\begin{figure}[t]
    \centering
    \includegraphics[width=\linewidth]{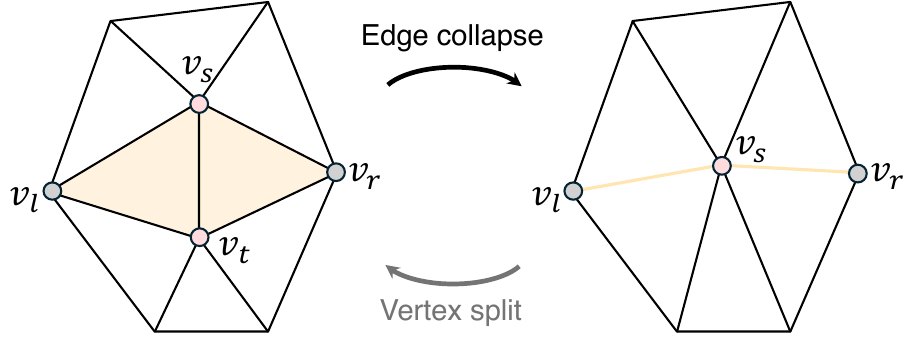}
    \caption{Illustration of edge collapse and its inverse operation, \ie vertex split. During edge collapse, vertex $v_t$ is collapsed into $v_s$, resulting in two degenerate triangles (shaded in yellow), which forms two new edges in the result mesh.}
    \label{fig:ecol}
\end{figure}

\subsection{Progressive Meshes Overview}

A \textit{progressive mesh}~\cite{hoppe1996progressive} (PM) proposes an efficient and lossless continuous-resolution representation for arbitrary triangle meshes. The representation is built off two observations: (i) that a single mesh transformation, \textit{edge collapse}, is sufficient for effective simplification of meshes; (ii) that edge collapse transformations are invertible via a \textit{vertex split} operation. In the following, we provide details of the edge collapse operation and its inversion with vertex splits, each illustrated in \cref{fig:ecol}.

The edge collapse operation, $\mathrm{ecol}({v_s, v_t})$, unifies two adjacent vertices $v_s$ and $v_t$ into a single vertex $v_s$. The operation results in the vanishing of two faces, $\{v_t, v_s, v_l\}$ and $\{v_s, v_t, v_r\}$, as well as the vertex $v_t$. In the general case, a new position for $v_s$ is also specified; however, for the case of \textit{half-edge collapse}, the original position for $v_s$ is kept.

An initial mesh, $\mathcal{M}$, can be simplified into a more coarse mesh, $\mathcal{M}_n$, by $n$ successive edge collapse operations:
$$\mathcal{M}_0 \xleftarrow[]{\mathrm{ecol}_{0}} \mathcal{M}_1 \xleftarrow[]{\mathrm{ecol}_{1}} \quad \cdots \quad \xleftarrow[]{\mathrm{ecol}_{n-1}} \mathcal{M}$$

The order in which the edge collapse operations are performed is determined such that each successive edge collapse results in the minimum increase in Quadratic Error Metrics (QEM)~\cite{garland1997surface} with respect to the original mesh. This process can be repeated until there are no candidate edge collapses, \eg without flipping face normals.

A vertex split, $\mathrm{vsplit}(v_s, v_l, v_r, v_t)$, defines the inverse of edge collapse. It restores $v_t$ and the two faces, $\{v_t, v_s, v_l\}$ and $\{v_s, v_t, v_r\}$, which vanished during edge collapse, as well as $v_s$ to its original position. In case $(v_s, v_t)$ is a boundary edge, either $v_l$ or $v_r$ will not exist, and there is only one face restored during the vertex split.

The combination of edge collapse and its inverse vertex splits enables the progressive mesh representation, $\mathrm{PM}(\mathcal{M})=(\mathcal{M}_0, \{\mathrm{vsplit}_0, \cdots, \mathrm{vsplit}_{n-1}\})$ to express an arbitrary triangle mesh, $\mathcal{M}$, as a combination of a coarse mesh, $\mathcal{M}_0$, obtained through edge collapse, and the sequence of $n$ vertex split records required to reverse them:
$$\mathcal{M}_0 \xrightarrow[]{\mathrm{vsplit}_{0}} \mathcal{M}_1 \xrightarrow[]{\mathrm{vsplit}_{1}}  \quad \cdots \quad \xrightarrow[]{\mathrm{vsplit}_{n-1}} \mathcal{M}$$

\subsection{\OURS{}}

Traditional progressive meshes require starting an initial detailed mesh to be simplified into a coarse mesh, $\mathcal{M}_0$, and recording each of these collapse steps as vertex split records to form the representation. We observe that the formation of the PM representation serves effectively as forward (edge collapse) and reverse (vertex split) processes that enable transition between detailed and coarse meshes. 

\OURS{} frames the creation of a detailed mesh as the generation of a PM representation. Analogous to denoising for diffusion models, we train a generative model to reverse edge collapse. Concretely, \OURS{} first generates a coarse mesh, $\mathcal{M}_0$, from scratch, before increasing the level of detail through the generation of vertex split records.

\subsubsection{Progressive Mesh Parameterization}\label{sec:pm_parameterization}

We frame generation as an autoregressive sequential modeling task, with the full mesh sequence taking the form:
\begin{gather*}
\resizebox{\linewidth}{!}{
\small
$
\begin{aligned}
\texttt{M}: \; [\,&\texttt{<bos>}, \quad \texttt{[M\_0 sequence]}, \quad \texttt{<sep>}, 
&& \#\mathcal{M}_0 \\
 &\texttt{[vsplit\_0]}, \quad \texttt{...}, \quad \texttt{[vsplit\_n-1]}, \quad \texttt{<eos>}\,]&& \#\mathrm{vsplits}\\
\end{aligned}
$
}
\end{gather*}

In the following, we will define the formation of the subsequences \texttt{[M\_0 sequence]} and \texttt{[vsplit]}.

\paragraph{$\mathcal{M}_0$: Coarse Mesh Tokenization.}
For the initial coarse mesh, we follow the tokenization scheme defined in MeshXL~\cite{chen2024meshxl}. In this formulation, embeddings are learned for discretized coordinates in an $N^3$ grid. A vertex is represented via the sequential look-up of $x$-value, $y$-value, and $z$-value. A face is then constructed as the concatenation of its 3 vertices, totaling 9 tokens. The full mesh sequence is then defined as the concatenation of its constituent faces.
\bgroup
\small
\begin{gather*}
\begin{aligned}
\texttt{v}:\; & [\,\texttt{<x>},\,\texttt{<y>},\,\texttt{<z>}\,], &&\#\mathrm{vertex} \\
\texttt{F}:\; & [\,\texttt{[v\_1]},\,\texttt{[v\_2]},\,\texttt{[v\_3]}\,], \quad &&\#\mathrm{face} \\
\texttt{M}:\; & [\,\texttt{[F\_1]},\,\texttt{[F\_2]},\,\texttt{...},\,\texttt{[F\_N]}\,] \quad &&\#\mathrm{mesh}
\end{aligned}
\end{gather*}
\egroup

Following \cite{siddiqui2024meshgpt}, we sort vertices in $z$-$y$-$x$ order (lower to higher). Within each face, vertices are cylindrically permuted to have the lowest-indexed vertex be first.

\begin{figure}[t]
    \centering
    \includegraphics[width=.8\linewidth,trim=0 0.5em 0 0,clip]{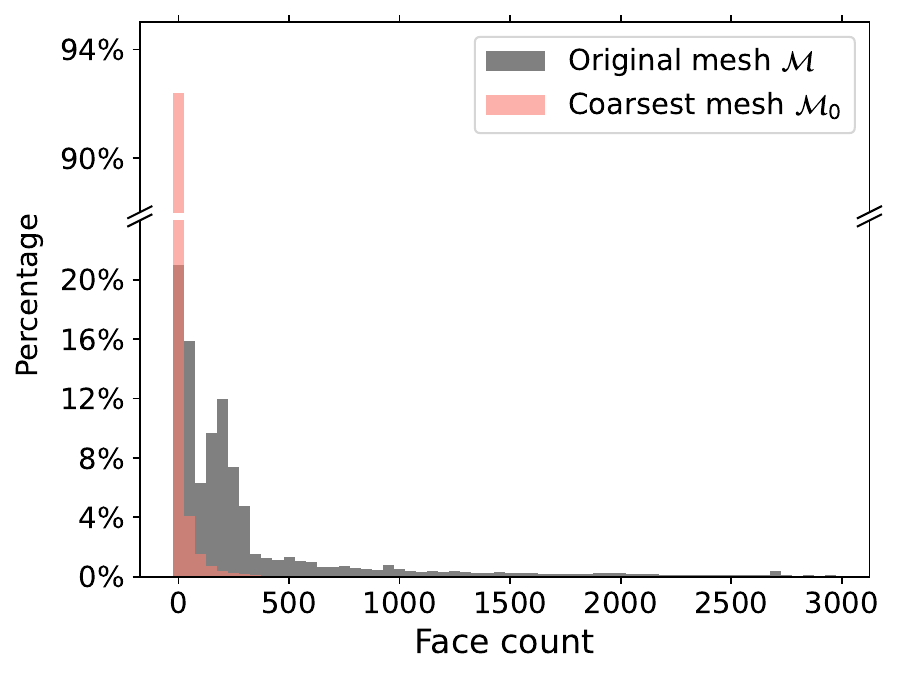}
    \caption{Face count distribution of the coarsest mesh $\mathcal{M}_0$ and original mesh $\mathcal{M}$. The coarsest meshes ($\mathcal{M}_0$) contain significantly fewer faces than original ($\mathcal{M}$), with an average of 18 and 457 faces, respectively.}
    \label{fig:distrbution_num_faces_m_t_m_0}
\end{figure}

Although MeshXL~\cite{chen2024meshxl} leverages this tokenization scheme to produce full detailed meshes, in \OURS{} only the coarsest level of the mesh is parameterized in this way. These coarse meshes consist of substantially fewer faces than the full detailed mesh, as demonstrated in \cref{fig:distrbution_num_faces_m_t_m_0}, and the resulting coarse mesh only accounts for 5.68\% of the total sequence length on average.

\paragraph{Vertex Split Generation.} 

Consider the $k$-th vertex split operation $\mathrm{vsplit}_k$, which converts a lower-detailed mesh $\mathcal{M}_k$ into a higher-detailed mesh $\mathcal{M}_{k+1}$. We denote the neighbor vertices of vertex $v$ in $\mathcal{M}_k$ as $N_k(v)$. After determining the target vertex $v_s$ to split (in $\mathcal{M}_k$) and the new vertex $v_t$ (in $\mathcal{M}_{k+1}$), we need to obtain $N_{k+1}(v_s)$ and $N_{k+1}(v_t)$ from the vertex split record to connect $v_s$ and $v_t$ with correct neighbor vertices in $\mathcal{M}_{k+1}$, which are prohibitive to generate as the number of neighbor vertices $|N_{k+1}(v_s)|$ and $|N_{k+1}(v_t)|$ is often large.

However, as mesh $\mathcal{M}_{k}$ is generated from $\mathcal{M}_{k+1}$ during edge collapse by merging $v_t$ into $v_s$, we have
\begin{equation}
    N_{k+1} (v_s) \cap N_{k+1}(v_t) = \{ v_l, v_r\} \label{eqn:vertex_neighbor_intersect}
\end{equation}
\begin{equation}
    N_{k+1} (v_s) \cup N_{k+1}(v_t) - \{ v_s, v_t \} = N_k(v_s) \label{eqn:vertex_neighbor_union}
\end{equation}
which indicates we will only need to record vertex $v_l$ and $v_r$. They split the vertex ring surrounding $v_s$ in $\mathcal{M}_k$, where two halves of the vertices on the ring (top and bottom vertices in \cref{fig:ecol}) belong to $v_t$ and $v_s$ respectively in $\mathcal{M}_{k+1}$. However, an ambiguity arises in determining which half of the ring corresponds to $v_t$'s neighbors in $\mathcal{M}_{k+1}$. In the following, we demonstrate how the half-edge data structure \cite{weiler1986topological} can be utilized to resolve this ambiguity.

\begin{figure}[t]
    \centering
    \includegraphics[width=\linewidth]{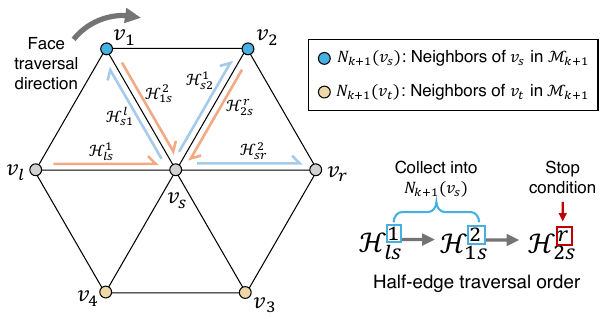}
    \caption{Illustration of half-edge data structure and the traversal process to determine the neighbors of $v_s$ and $v_t$ in $\mathcal{M}_{k+1}$. Starting from half-edge $\mathcal{H}_{ls}^1$, we traverse the faces in clockwise direction until $\mathcal{H}_{sr}^2$, where we collect the vertices $v_1$ and $v_2$ into $N_{k+1}(v_s)$.}
    \label{fig:half-edge-traversal}
\end{figure}

Denote $\mathcal{H}_{ij}^k$ as a half-edge pointing from $v_i$ to $v_j$, with $v_k$ being the third vertex in the associated face. As shown in \cref{fig:half-edge-traversal}, we begin the traversal from $\mathcal{H}_{ls}^1$ and proceed to the \textit{twin} of the \textit{next} half-edge, $\mathcal{H}_{1s}^2$. This operation is repeated until we reach $\mathcal{H}_{\boldsymbol{\cdot} s}^r$. Throughout this process, we traverse the faces above $(v_l, v_s)$ and $(v_s, v_r)$ clockwise, obtaining:
\begin{equation}
    \{ v_k \, | \, \mathcal{H}_{\boldsymbol{\cdot} s}^k, \, k\ne r \} = N_{k+1}(v_s) - \{v_l, v_r, v_t\}
\end{equation}
where $\mathcal{H}_{\boldsymbol{\cdot} s}^k$ is the half-edge encountered during traversal.

In conjunction with \cref{eqn:vertex_neighbor_intersect,eqn:vertex_neighbor_union}, $N_{k+1}(v_t)$ can also be determined. When $(v_s, v_t)$ is a boundary in $\mathcal{M}_{k+1}$, we follow either $\mathcal{H}_{ls}^{\boldsymbol{\cdot}}$ clockwise or $\mathcal{H}_{sr}^{\boldsymbol{\cdot}}$ counterclockwise until we reach a half-edge of the boundary.

\begin{figure}[t]
    \centering
    \includegraphics[width=\linewidth]{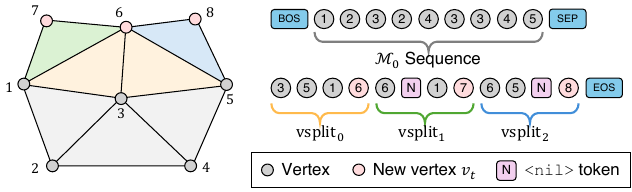}
    \caption{Illustration of \OURS{} tokenization. The sequence begins with base mesh $\mathcal{M}_0$, followed by vertex split subsequences. A special \texttt{<nil>} token indicates either $v_l$ or $v_r$ does not exist.}
    \label{fig:tokenization}
\end{figure}

After identifying the neighbors of $v_s$ and $v_t$ in the mesh $\mathcal{M}_{k+1}$, we add the new vertex $v_t$ to the mesh $\mathcal{M}_k$ and reconnect the vertices in $N_{k+1}(v_t)$ to $v_t$. Finally, we restore two faces, $\{v_s, v_l, v_t\}$ and $\{ v_r, v_s, v_t \}$, which are associated with the half-edges $\mathcal{H}_{sl}^t$ and $\mathcal{H}_{rs}^t$, respectively. Using the half-edge data structure, the orientation of the newly created faces remains consistent with the rest of the mesh. When $(v_s, v_t)$ lies on the boundary, only one face is restored.

Hence, with the above half-edge formulation, each vertex split operation can be uniquely defined by the selection of the target vertex $v_s$ and two neighbors $v_l$ and $v_r$, as well as the position to place the new vertex $v_t$.

\begin{figure*}[ht]
    \centering
    \includegraphics[width=\linewidth,trim=0 0 0 6pt,clip]{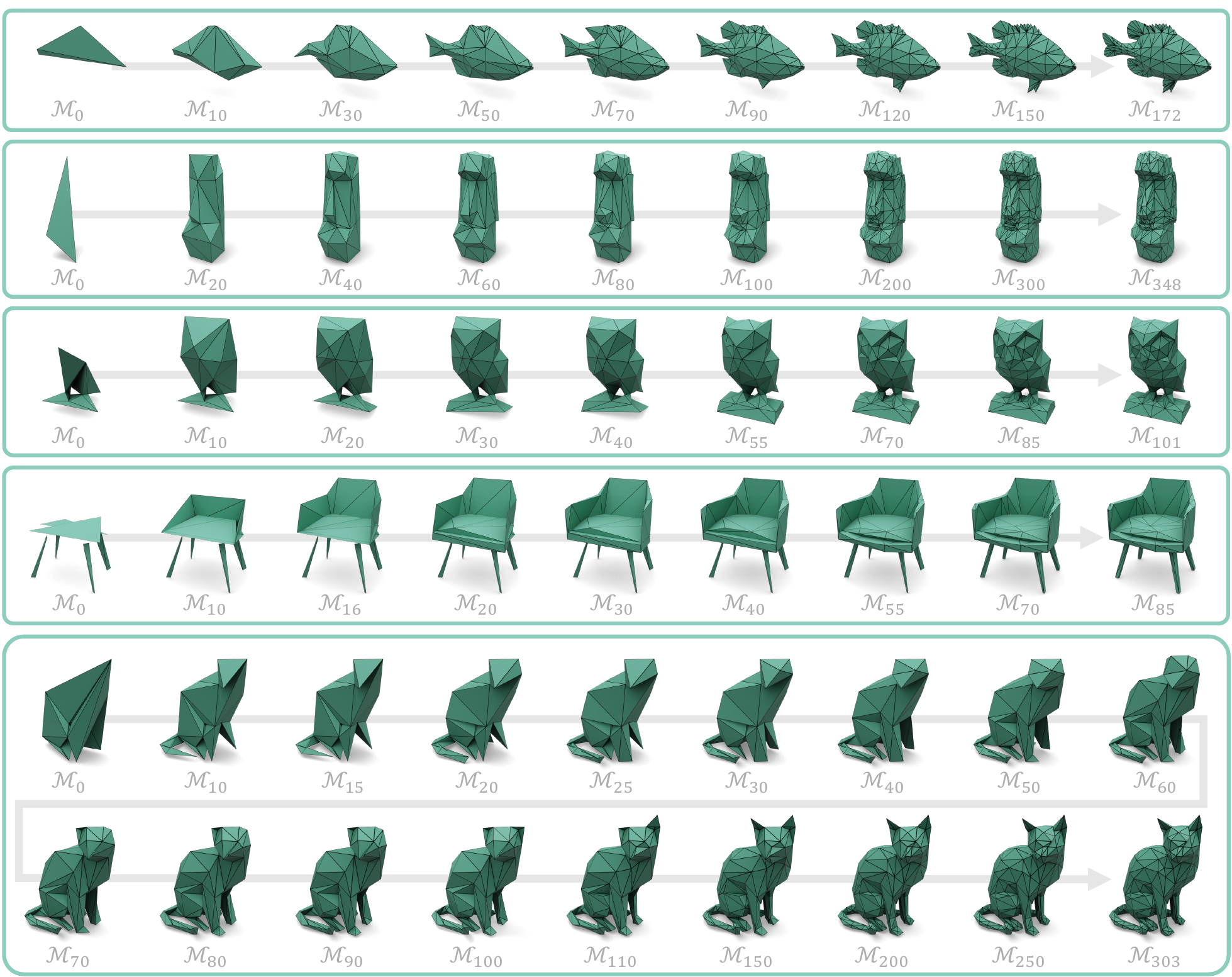}
    \caption{Generation process for \OURS{}. $\mathcal{M}_0$ represents the predicted initial coarsest mesh, followed by subsequent meshes generated through the predicted vertex split sequence.}
    \label{fig:unconditional}
\end{figure*}

\paragraph{Vertex Split Tokenization.} Although $v_s$, $v_l$ and $v_r$ only require references to existing vertices in the mesh, in practice we implement this reference as a raw prediction of each vertex to avoid a vocabulary size proportional to the sequence length. Thus, in the majority of cases, each vertex split is represented by a subsequence of 12 tokens (or 10 tokens when $(v_s, v_t)$ is a boundary edge):
\bgroup
\small
\begin{gather*}
\begin{aligned}
\texttt{vsplit}: [\, &\texttt{<s\_x>},\,\texttt{<s\_y>},\,\texttt{<s\_z>}, \quad &&\#v_s \\
& \texttt{<l\_x>},\,\texttt{<l\_y>},\,\texttt{<l\_z>} \,|\,\texttt{<nil>}, \quad &&\#v_l \\
& \texttt{<r\_x>},\,\texttt{<r\_y>},\,\texttt{<r\_z>} \,|\,\texttt{<nil>}, \quad &&\#v_r \\
& \texttt{<t\_x>},\,\texttt{<t\_y>},\,\texttt{<t\_z>} \,] \quad &&\#v_t
\end{aligned}
\end{gather*}
\egroup
where the special token \texttt{<nil>} signifies $v_l$ or $v_r$ does not exist, and only one of $v_l$ and $v_r$ is allowed to be \texttt{<nil>} for a valid vertex split subsequence. In \cref{fig:tokenization} we illustrate the entire tokenization process.

The above progressive mesh serialization can then be trained with the standard next-token prediction target.

\paragraph{Vertex Split Decoding.} The vertex split sequence must be consistent with the actual geometry to be valid. We maintain a state machine and perform vertex split on the fly as new tokens are generated. After the coarsest mesh $\mathcal{M}_0$ is generated, we initialize the state. During each generation step, we enforce $v_s$ to be a vertex in the current mesh $\mathcal{M}_k$, $(v_s, v_l)$, $(v_s, v_r)$ are valid edges (if $v_l$ and $v_r$ are not \texttt{<nil>}), and only one of $v_l$ and $v_r$ can be \texttt{<nil>}. Lastly, we decode $v_t$, and perform vertex split with generated $v_s, v_l, v_r, v_t$ information.

\subsection{Conditional Generation}

We consider generation conditioned on shapes. We adopt a pre-trained point-cloud encoder~\cite{zhao2023michelangelo}. A LLaVa-style~\cite{liu2023visual} projector is leveraged to project the condition features to the token embedding space. The projected feature tokens are prepended to the mesh sequence as the prefix. We supervise training with the next-token prediction objective while masking the loss for the prefix tokens.

%% file: sec/4_experiments.tex
\textbf{}\begin{table*}[t]
    \centering
    \begin{tabular}{llcccc} \toprule
        \multirow{2}{*}{Method} & \multirow{2}{*}{Tokenization} & COV  & MMD & 1-NNA  & JSD \\
         & & (\%, $\uparrow$) & ($\times10^3$, $\downarrow$) & (\%) & ($\downarrow$) \\
        \midrule
        MeshXL~\cite{chen2024meshxl} & Flattened Coords. & \textbf{51.76} & 8.30 & 50.84 & 3.81 \\
        MeshAnything V2~\cite{chen2024meshanythingv2} & AMT & 50.33 & 8.50 & 52.25 & 4.84 \\
        EdgeRunner~\cite{tang2024edgerunner} & EdgeBreaker~\cite{rossignac2002edgebreaker} & \underline{51.39} & \textbf{7.81} & \underline{49.44} & \underline{3.22}\\
        \OURS~(Ours) & Progressive & 51.03 & \underline{8.29} & \textbf{50.22} & \textbf{2.89} \\\bottomrule
    \end{tabular}
    \caption{Quantitative comparisons with state-of-the-art methods for unconditional mesh generation. Best results are \textbf{bolded}, second best are \underline{underlined}. \OURS{} can generate meshes with comparable quality while enjoying the benefits of continuous level of detail.}
    \label{tab:unconditional}
\end{table*}

\begin{figure*}[t]
    \centering
    \includegraphics[width=\linewidth,trim=1em 1em 1em 0,clip]{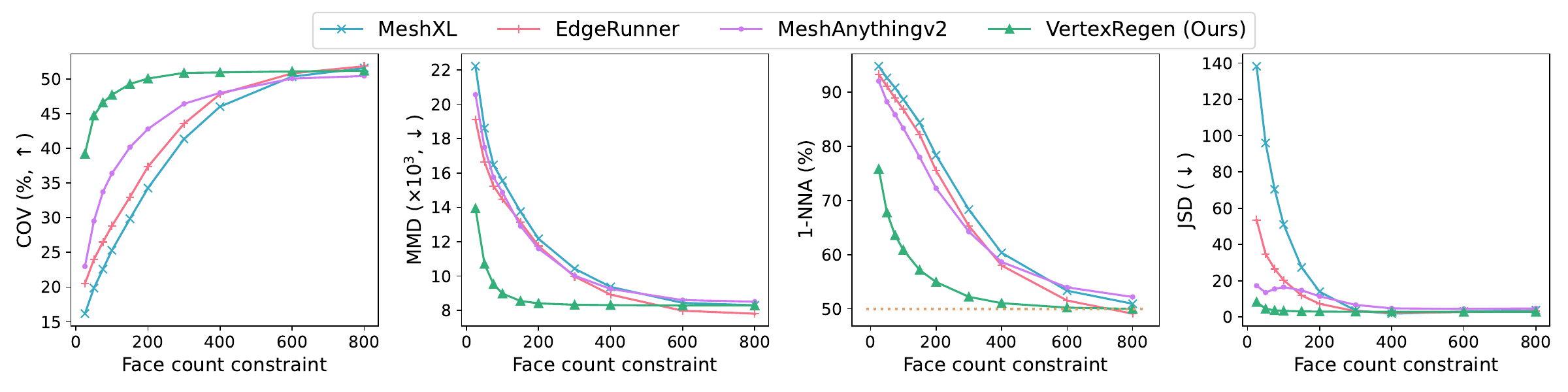}
    \caption{Unconditional generation under face count constraints. \OURS{} achieves significantly better COV, MMD, and 1-NNA in early stages of generation.
    }
    \label{fig:ablation_face_count}
\end{figure*}

\section{Experiments}

\subsection{Datasets}
We pre-train our model and the baselines presented using two primary sources of meshes: Objaverse-XL\footnote{We neither used assets from Sketchfab nor obtained any from the Polycam website.} \cite{objaverseXL} and an additional set of licensed artist-created meshes. Initially, we select meshes containing fewer than 8,000 faces without applying decimation. We use the CGAL \cite{fabri2009cgal} library to implement edge collapse and vertex split operations, where we modify the vertex placement to either $v_s$ or $v_t$. We filter out non-manifold meshes and those that cannot be processed, which results in a final dataset of approximately 1.5M meshes with an average of 457 faces. For unconditional generation evaluation, we construct a high-quality subset by further filtering with the alignment split of Objaverse-XL and with fewer than 800 faces. This yields approximately 18k meshes. We adopt a similar filtering process for shape-conditioned experiments.

We apply robust data augmentations to input meshes, including random shift within range $[-0.1, 0.1]$, random scaling between $[0.9, 1.1]$, random rotation by 0°, 90°, 180°, or 270°. During pre-training, we handle sequences exceeding the context window by discarding vertex split subsequences, while for other baselines, we truncate the tokenized sequence directly. The set for quantitative evaluation is chosen to avoid truncation for any method evaluated.

\subsection{Implementation Details}

\OURS{} and all other baselines are built upon the pre-trained OPT-350M \cite{zhang2022opt}, with newly initialized token embeddings and position embeddings. We train our model on 64 H100 GPUs for approximately four days, with an effective batch size of 256. We use the AdamW \cite{loshchilov2017decoupled} optimizer with a weight decay of 0.1, and betas $(0.9, 0.95)$. A cosine scheduler is employed, starting with an initial learning rate of $1 \times 10^4$ and gradually decreasing to $5 \times 10^6$. We clip the gradient norm to 1.0. During inference, we adopt the top-$p$ sampling strategy with $p=0.95$ by default.

\begin{figure*}[ht]
    \centering
    \includegraphics[width=\linewidth]{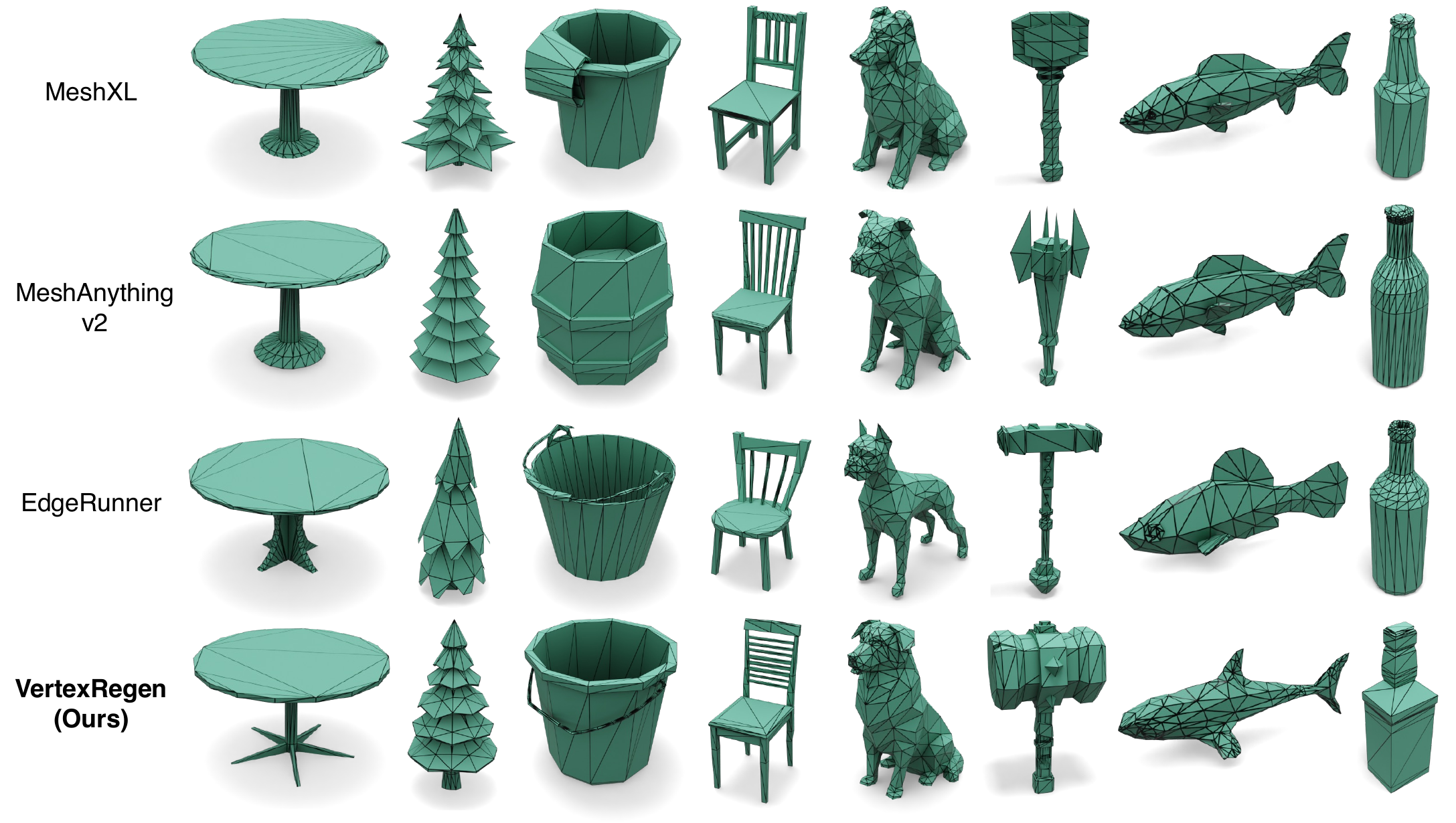}
    \caption{Qualitative comparison with state-of-the-art methods. \OURS{} is able to generate meshes with comparable quality to other baselines.}
    \label{fig:uncond_comparison}
\end{figure*}

\begin{figure*}[ht]
    \centering
    \includegraphics[width=\linewidth]{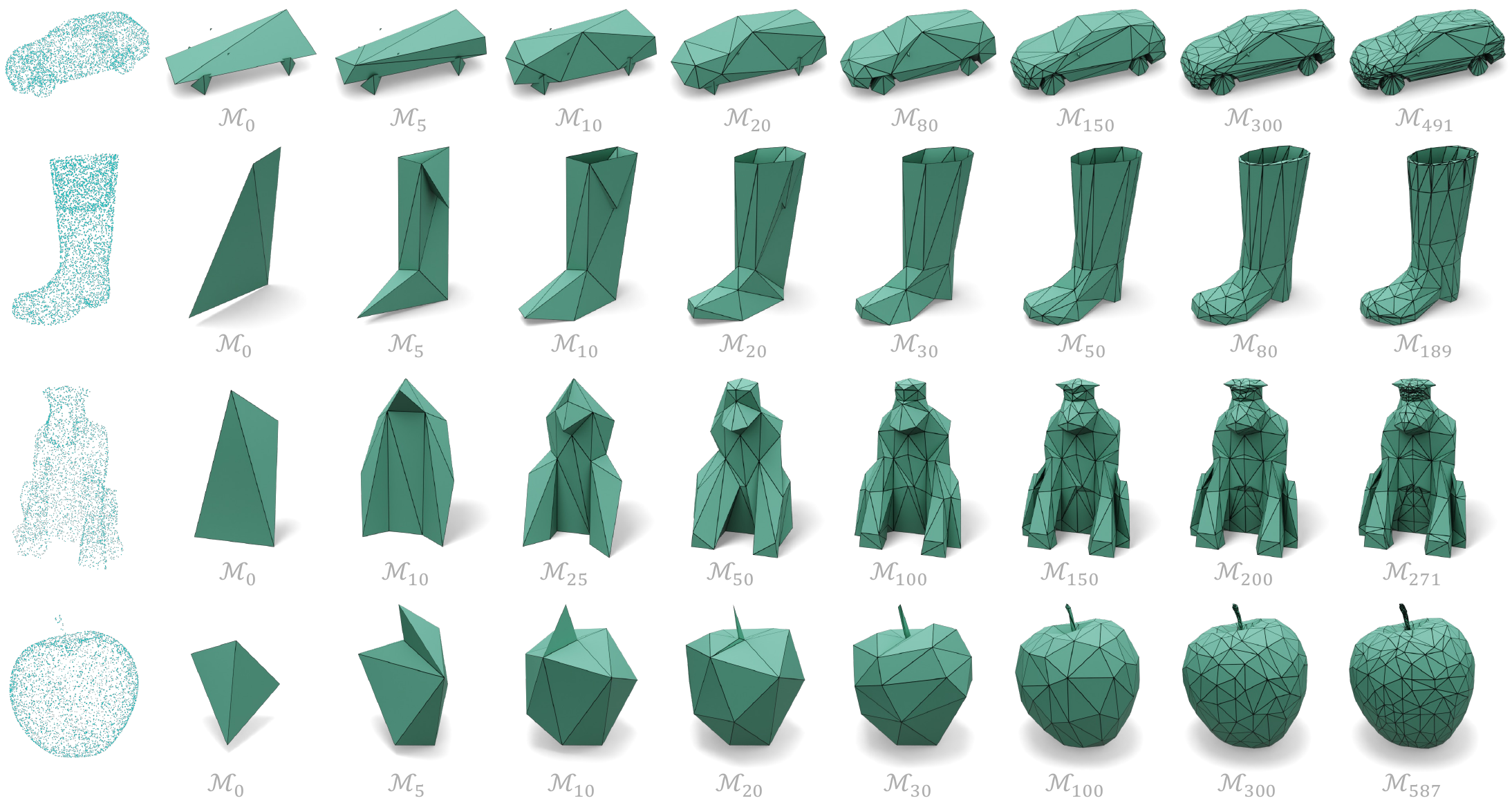}
    \caption{Qualitative results on shape-conditioned generation on meshes held out from training. The first column shows the point cloud used as the condition, followed by the generation sequence progressing from the coarsest mesh $\mathcal{M}_0$ to the final output in the last column.}
    \label{fig:conditional}
\end{figure*}

\subsection{Results}

\subsubsection{Unconditional Generation}

We follow evaluation protocols from prior works \cite{siddiqui2024meshgpt,chen2024meshxl,alliegro2023polydiff}, employing point-cloud-based metrics to assess unconditional generation. Specifically, we sample the same number of meshes as the evaluation dataset and randomly sample 2,048 points per mesh. Coverage (COV) measures the diversity of the generated samples, where higher values indicate greater diversity. Minimum Matching Distance (MMD) computes the average distance from each reference sample to its nearest neighbor in the generated set, serving as a measure of generation quality, with lower values being preferable. 1-Nearest Neighbor Accuracy (1-NNA) evaluates both diversity and quality, where an optimal value is achieved at 50\%. Additionally, we compute Jensen-Shannon Divergence (JSD) to directly quantify the similarity between the generated and reference distributions. We present results in \cref{tab:unconditional}. Our method can generate meshes of comparable quality to other state-of-the-art methods, with the added advantage of a continuous level of detail. We also show qualitative examples and their generation process in \cref{fig:unconditional}, and more generation results comparing with state-of-the-art methods in \cref{fig:uncond_comparison}.

\paragraph{Generation with a face count constraint.} We compare our method with the baselines under varying face count constraints. Our approach allows for generation to be paused at any point, accommodating the varying constraints. In contrast, to adapt to these constraints, baseline methods require either: (i) directly truncating the generation process, yielding incomplete intermediate meshes, or (ii) fine-tuning with an additional face count condition token.

For the direct truncation version, we evaluate unconditional generation metrics \wrt different face count constraints and present the results in \cref{fig:ablation_face_count}. Additionally, we adopt a similar approach as proposed in EdgeRunner, where we split $[1, 800]$ range into 4 buckets and prepend the conditioning token to the mesh sequence. We compare \OURS{} and MeshXL fine-tuned with such conditioning scheme at face constraint 400, as shown in \cref{tab:face_count}.

Due to its ability to generate meshes with a continuous level of detail from coarse to fine, \OURS{} effectively captures the overall structure even with very limited face counts, yielding significantly better COV, MMD, and 1-NNA early in the process. As the face limit increases, all methods demonstrate improved results. Notably, for baselines such as EdgeRunner, while they enhance tokenization efficiency over MeshXL, they do not fundamentally address continuous level-of-detail generation and thus follow a similar trend to other baselines in the plot.

\subsubsection{Conditional Generation}

We show qualitative examples of shape-conditioned generation in \cref{fig:conditional}. We condition on 4,096 sampled points with normals from dense meshes. \OURS{} can generate a coarse $\mathcal{M}_0$, starting as simply as a tetrahedron, and progressively refine it by generating a sequence of vertex splits.

\subsection{Ablation Studies}

\paragraph{Tokenization efficiency.} We compute the tokenized sequence length of our proposed tokenization scheme (\cref{sec:pm_parameterization}) on our dataset and report the average compression ratio relative to MeshXL (9 tokens per face) in \cref{tab:compress_ratio}. For \OURS{}, there are two primary sequence types: (i) a MeshXL-style sequence for the initial coarsest mesh $\mathcal{M}_0$ and (ii) sequences encoding vertex splits. The latter requires 12 tokens for two non-boundary faces or 10 tokens per boundary face. As a result, the highest compression is achieved when $\mathcal{M}_0$ is minimized and all vertex splits occur on non-boundary faces. On average, \OURS{} achieves a compression ratio of 0.73, approaching the theoretical limit of 0.67. The discrepancy arises from the overhead introduced by $\mathcal{M}_0$ (5.68\% of all total tokenized length) and boundary vertex splits (3.64\% of all vertex splits).

Without the half-edge data structure, identifying which half of the ring surrounding $v_s$ is associated with $v_s$ or $v_t$ in $\mathcal{M}_{k+1}$ requires recording an additional vertex. This increases the token count to 15 per two non-boundary faces (or 13 per boundary face) and results in an average increase of 22\% in tokenized sequence length.

\begin{table}[t]
\centering
\resizebox{\linewidth}{!}{
\setlength{\tabcolsep}{4pt}
\begin{tabular}{lcccccc} \toprule
\multirow{2}{*}{@400 Faces} & COV  & MMD & 1-NNA  & JSD & \multirow{2}{*}{$\vert F\vert$} & \multirow{2}{*}{$\vert V\vert$} \\
& (\%, $\uparrow$) & ($\times10^3$, $\downarrow$) & (\%) & ($\downarrow$) \\
\midrule
\OURS{} & \textbf{50.92} & \textbf{8.31} & \textbf{51.03} & \textbf{2.88} & 264 & 147 \\
MeshXL (w/ FCC) & 41.20 & 10.03 & 59.06 & 5.19 & 308 & 168 \\ \bottomrule
\end{tabular}
}
\caption{Comparison between \OURS{} and MeshXL with face count condition (FCC) when face count constraint is 400.}
\label{tab:face_count}
\end{table}

\begin{table}[t]
\centering
\resizebox{\linewidth}{!}{
\setlength{\tabcolsep}{4pt}
\begin{tabular}{l c c c c c}
\toprule
 &  \multirow{2}{*}{MeshXL} & MeshAnything & \multirow{2}{*}{EdgeRunner} & \multicolumn{2}{c}{\OURS{}} \\
 & & v2 & & w/ HE & w/o HE \\
\midrule
Compression  &  1.0 & 0.46 & 0.47 & 0.73 & 0.89 \\
\bottomrule
\end{tabular}
}
\caption{Compression ratio of different tokenization schemes, \OURS{} with and without leveraging a half-edge (HE) structure.}
\label{tab:compress_ratio}
\end{table}

\paragraph{Guided decoding.} In \cref{tab:ablation_guided_decoding}, we ablate the effects of geometry-constrained decoding. As one step of the vertex split operation is dependent on all the preceding operations, predicting an invalid vertex split may break the chain and end the generation prematurely. With guided decoding, the model is able to generate longer sequences with more faces.

\begin{table}[t]
\centering
\resizebox{\linewidth}{!}{

\begin{tabular}{lcccccc} \toprule
    Guided & COV  & MMD & 1-NNA  & JSD & \multirow{2}{*}{$\vert F\vert$} & \multirow{2}{*}{$\vert V\vert$} \\
    Decoding & (\%, $\uparrow$) & ($\times10^3$, $\downarrow$) & (\%) & ($\downarrow$) \\
    \midrule
    w/o &  \textbf{51.12} & 8.31 & 50.75 & 3.37 & 211 & 120 \\
    w/ & 51.03 & \textbf{8.29} & \textbf{50.22} & \textbf{2.89} & 320 & 176 \\\bottomrule
\end{tabular}

}
\caption{Effects on geometry-guided decoding.}
\label{tab:ablation_guided_decoding}
\end{table}

%% file: sec/5_conclusion.tex
\section{Conclusion}

In this work, we introduced \OURS{}, a novel mesh generation framework that enables continuous levels of detail through a generative process based on vertex splits. Unlike conventional auto-regressive approaches that synthesize meshes in a partial-to-complete manner, \OURS{} reinterprets mesh generation as the reversal of edge collapse, providing an effectively \textit{anytime} solution to mesh generation. Our experimental results demonstrate that \OURS{} achieves competitive performance compared to state-of-the-art methods while offering the unique advantage of halting generation at any stage to obtain meshes at different levels of detail.

%% file: main.bbl
\begin{thebibliography}{59}
\providecommand{\natexlab}[1]{#1}
\providecommand{\url}[1]{\texttt{#1}}
\expandafter\ifx\csname urlstyle\endcsname\relax
  \providecommand{\doi}[1]{doi: #1}\else
  \providecommand{\doi}{doi: \begingroup \urlstyle{rm}\Url}\fi

\bibitem[Alliegro et~al.(2023)Alliegro, Siddiqui, Tommasi, and Nie{\ss}ner]{alliegro2023polydiff}
Antonio Alliegro, Yawar Siddiqui, Tatiana Tommasi, and Matthias Nie{\ss}ner.
\newblock Polydiff: Generating 3d polygonal meshes with diffusion models.
\newblock \emph{arXiv preprint arXiv:2312.11417}, 2023.

\bibitem[Chen et~al.(2024{\natexlab{a}})Chen, Chen, Pang, Zeng, Cheng, Fu, Yin, Wang, Yu, Yu, et~al.]{chen2024meshxl}
Sijin Chen, Xin Chen, Anqi Pang, Xianfang Zeng, Wei Cheng, Yijun Fu, Fukun Yin, Billzb Wang, Jingyi Yu, Gang Yu, et~al.
\newblock Meshxl: Neural coordinate field for generative 3d foundation models.
\newblock \emph{Advances in Neural Information Processing Systems}, 37:\penalty0 97141--97166, 2024{\natexlab{a}}.

\bibitem[Chen et~al.(2024{\natexlab{b}})Chen, Wang, Luo, Wang, Chen, Zhu, Zhang, and Lin]{chen2024meshanythingv2}
Yiwen Chen, Yikai Wang, Yihao Luo, Zhengyi Wang, Zilong Chen, Jun Zhu, Chi Zhang, and Guosheng Lin.
\newblock Meshanything v2: Artist-created mesh generation with adjacent mesh tokenization.
\newblock \emph{arXiv preprint arXiv:2408.02555}, 2024{\natexlab{b}}.

\bibitem[Chen et~al.(2025)Chen, He, Huang, Ye, Chen, Tang, Cai, Yang, Yu, Lin, and Zhang]{chen2024meshanything}
Yiwen Chen, Tong He, Di Huang, Weicai Ye, Sijin Chen, Jiaxiang Tang, Zhongang Cai, Lei Yang, Gang Yu, Guosheng Lin, and Chi Zhang.
\newblock Meshanything: Artist-created mesh generation with autoregressive transformers.
\newblock In \emph{The Thirteenth International Conference on Learning Representations}, 2025.

\bibitem[Chen et~al.(2023)Chen, Kim, Aigerman, and Jacobson]{chen2023neural}
Yun-Chun Chen, Vladimir Kim, Noam Aigerman, and Alec Jacobson.
\newblock Neural progressive meshes.
\newblock In \emph{ACM SIGGRAPH 2023 Conference Proceedings}, pages 1--9, 2023.

\bibitem[Chen et~al.(2020)Chen, Tagliasacchi, and Zhang]{chen2020bsp}
Zhiqin Chen, Andrea Tagliasacchi, and Hao Zhang.
\newblock Bsp-net: Generating compact meshes via binary space partitioning.
\newblock In \emph{Proceedings of the IEEE/CVF Conference on Computer Vision and Pattern Recognition (CVPR)}, 2020.

\bibitem[Cheng et~al.(2023)Cheng, Lee, Tulyakov, Schwing, and Gui]{cheng2023sdfusion}
Yen-Chi Cheng, Hsin-Ying Lee, Sergey Tulyakov, Alexander~G. Schwing, and Liang-Yan Gui.
\newblock Sdfusion: Multimodal 3d shape completion, reconstruction, and generation.
\newblock In \emph{Proceedings of the IEEE/CVF Conference on Computer Vision and Pattern Recognition (CVPR)}, pages 4456--4465, 2023.

\bibitem[Dai and Niessner(2019)]{dai2019scan2mesh}
Angela Dai and Matthias Niessner.
\newblock Scan2mesh: From unstructured range scans to 3d meshes.
\newblock In \emph{Proceedings of the IEEE/CVF Conference on Computer Vision and Pattern Recognition (CVPR)}, 2019.

\bibitem[Deitke et~al.(2023)Deitke, Liu, Wallingford, Ngo, Michel, Kusupati, Fan, Laforte, Voleti, Gadre, et~al.]{objaverseXL}
Matt Deitke, Ruoshi Liu, Matthew Wallingford, Huong Ngo, Oscar Michel, Aditya Kusupati, Alan Fan, Christian Laforte, Vikram Voleti, Samir~Yitzhak Gadre, et~al.
\newblock Objaverse-xl: A universe of 10m+ 3d objects.
\newblock \emph{Advances in Neural Information Processing Systems}, 36:\penalty0 35799--35813, 2023.

\bibitem[Fabri and Pion(2009)]{fabri2009cgal}
Andreas Fabri and Sylvain Pion.
\newblock Cgal: the computational geometry algorithms library.
\newblock In \emph{Proceedings of the 17th ACM SIGSPATIAL International Conference on Advances in Geographic Information Systems}, page 538–539, New York, NY, USA, 2009. Association for Computing Machinery.

\bibitem[Gao et~al.(2022)Gao, Shen, Wang, Chen, Yin, Li, Litany, Gojcic, and Fidler]{gao2022get3d}
Jun Gao, Tianchang Shen, Zian Wang, Wenzheng Chen, Kangxue Yin, Daiqing Li, Or Litany, Zan Gojcic, and Sanja Fidler.
\newblock Get3d: A generative model of high quality 3d textured shapes learned from images.
\newblock In \emph{Advances In Neural Information Processing Systems}, 2022.

\bibitem[Garland and Heckbert(1997)]{garland1997surface}
Michael Garland and Paul~S. Heckbert.
\newblock Surface simplification using quadric error metrics.
\newblock In \emph{Proceedings of the 24th Annual Conference on Computer Graphics and Interactive Techniques}, page 209–216, USA, 1997. ACM Press/Addison-Wesley Publishing Co.

\bibitem[Groueix et~al.(2018)Groueix, Fisher, Kim, Russell, and Aubry]{groueix2018papier}
Thibault Groueix, Matthew Fisher, Vladimir~G. Kim, Bryan~C. Russell, and Mathieu Aubry.
\newblock A papier-mâché approach to learning 3d surface generation.
\newblock In \emph{Proceedings of the IEEE Conference on Computer Vision and Pattern Recognition (CVPR)}, 2018.

\bibitem[Gu\'edon and Lepetit(2024)]{guedon2024sugar}
Antoine Gu\'edon and Vincent Lepetit.
\newblock Sugar: Surface-aligned gaussian splatting for efficient 3d mesh reconstruction and high-quality mesh rendering.
\newblock In \emph{Proceedings of the IEEE/CVF Conference on Computer Vision and Pattern Recognition (CVPR)}, pages 5354--5363, 2024.

\bibitem[Hao et~al.(2024)Hao, Romero, Lin, and Liu]{hao2024meshtron}
Zekun Hao, David~W Romero, Tsung-Yi Lin, and Ming-Yu Liu.
\newblock Meshtron: High-fidelity, artist-like 3d mesh generation at scale.
\newblock \emph{arXiv preprint arXiv:2412.09548}, 2024.

\bibitem[Hong et~al.(2024)Hong, Zhang, Gu, Bi, Zhou, Liu, Liu, Sunkavalli, Bui, and Tan]{hong2024lrm}
Yicong Hong, Kai Zhang, Jiuxiang Gu, Sai Bi, Yang Zhou, Difan Liu, Feng Liu, Kalyan Sunkavalli, Trung Bui, and Hao Tan.
\newblock {LRM}: Large reconstruction model for single image to 3d.
\newblock In \emph{The Twelfth International Conference on Learning Representations}, 2024.

\bibitem[Hoppe(1996)]{hoppe1996progressive}
Hugues Hoppe.
\newblock Progressive meshes.
\newblock In \emph{Proceedings of the 23rd Annual Conference on Computer Graphics and Interactive Techniques}, page 99–108, New York, NY, USA, 1996. Association for Computing Machinery.

\bibitem[Jun and Nichol(2023)]{jun2023shap}
Heewoo Jun and Alex Nichol.
\newblock Shap-e: Generating conditional 3d implicit functions.
\newblock \emph{arXiv preprint arXiv:2305.02463}, 2023.

\bibitem[Li et~al.(2023)Li, Duan, Zhou, and Lu]{li2023diffusion}
Muheng Li, Yueqi Duan, Jie Zhou, and Jiwen Lu.
\newblock Diffusion-sdf: Text-to-shape via voxelized diffusion.
\newblock In \emph{Proceedings of the IEEE/CVF Conference on Computer Vision and Pattern Recognition (CVPR)}, pages 12642--12651, 2023.

\bibitem[Lindstrom and Turk(1998)]{lindstrom1998fast}
Peter Lindstrom and Greg Turk.
\newblock Fast and memory efficient polygonal simplification.
\newblock In \emph{Proceedings Visualization'98 (Cat. No. 98CB36276)}, pages 279--286. IEEE, 1998.

\bibitem[Liu et~al.(2023)Liu, Li, Wu, and Lee]{liu2023visual}
Haotian Liu, Chunyuan Li, Qingyang Wu, and Yong~Jae Lee.
\newblock Visual instruction tuning.
\newblock In \emph{Advances in Neural Information Processing Systems}, pages 34892--34916. Curran Associates, Inc., 2023.

\bibitem[Liu et~al.(2024)Liu, Zeng, Wei, Shi, Chen, Xu, Zhang, Wang, Zhang, Liu, Wu, and Su]{liu2024meshformer}
Minghua Liu, Chong Zeng, Xinyue Wei, Ruoxi Shi, Linghao Chen, Chao Xu, Mengqi Zhang, Zhaoning Wang, Xiaoshuai Zhang, Isabella Liu, Hongzhi Wu, and Hao Su.
\newblock Meshformer: High-quality mesh generation with 3d-guided reconstruction model.
\newblock In \emph{Advances in Neural Information Processing Systems}, pages 59314--59341. Curran Associates, Inc., 2024.

\bibitem[Long et~al.(2024)Long, Guo, Lin, Liu, Dou, Liu, Ma, Zhang, Habermann, Theobalt, and Wang]{long2024wonder3d}
Xiaoxiao Long, Yuan-Chen Guo, Cheng Lin, Yuan Liu, Zhiyang Dou, Lingjie Liu, Yuexin Ma, Song-Hai Zhang, Marc Habermann, Christian Theobalt, and Wenping Wang.
\newblock Wonder3d: Single image to 3d using cross-domain diffusion.
\newblock In \emph{Proceedings of the IEEE/CVF Conference on Computer Vision and Pattern Recognition (CVPR)}, pages 9970--9980, 2024.

\bibitem[Lorensen and Cline(1987)]{lorensen1998marching}
William~E. Lorensen and Harvey~E. Cline.
\newblock Marching cubes: A high resolution 3d surface construction algorithm.
\newblock In \emph{Proceedings of the 14th Annual Conference on Computer Graphics and Interactive Techniques}, page 163–169, New York, NY, USA, 1987. Association for Computing Machinery.

\bibitem[Loshchilov and Hutter(2019)]{loshchilov2017decoupled}
Ilya Loshchilov and Frank Hutter.
\newblock Decoupled weight decay regularization.
\newblock In \emph{International Conference on Learning Representations}, 2019.

\bibitem[Luebke et~al.(2002)Luebke, Reddy, Cohen, Varshney, Watson, and Huebner]{luebke2002level}
David Luebke, Martin Reddy, Jonathan~D. Cohen, Amitabh Varshney, Benjamin Watson, and Robert Huebner.
\newblock \emph{Level of Detail for 3D Graphics}.
\newblock Morgan Kaufmann Publishers Inc., San Francisco, CA, USA, 2002.

\bibitem[Mescheder et~al.(2019)Mescheder, Oechsle, Niemeyer, Nowozin, and Geiger]{mescheder2019occupancy}
Lars Mescheder, Michael Oechsle, Michael Niemeyer, Sebastian Nowozin, and Andreas Geiger.
\newblock Occupancy networks: Learning 3d reconstruction in function space.
\newblock In \emph{Proceedings of the IEEE/CVF Conference on Computer Vision and Pattern Recognition (CVPR)}, 2019.

\bibitem[Mujkanovic et~al.(2024)Mujkanovic, Nsampi, Theobalt, Seidel, and Leimk{\"u}hler]{mujkanovic2024neural}
Felix Mujkanovic, Ntumba~Elie Nsampi, Christian Theobalt, Hans-Peter Seidel, and Thomas Leimk{\"u}hler.
\newblock Neural gaussian scale-space fields.
\newblock \emph{ACM Transactions on Graphics (TOG)}, 43\penalty0 (4):\penalty0 1--15, 2024.

\bibitem[Nash et~al.(2020)Nash, Ganin, Eslami, and Battaglia]{nash2020polygen}
Charlie Nash, Yaroslav Ganin, SM~Ali Eslami, and Peter Battaglia.
\newblock Polygen: An autoregressive generative model of 3d meshes.
\newblock In \emph{International conference on machine learning}, pages 7220--7229. PMLR, 2020.

\bibitem[Nichol et~al.(2022)Nichol, Jun, Dhariwal, Mishkin, and Chen]{nichol2022point}
Alex Nichol, Heewoo Jun, Prafulla Dhariwal, Pamela Mishkin, and Mark Chen.
\newblock Point-e: A system for generating 3d point clouds from complex prompts.
\newblock \emph{arXiv preprint arXiv:2212.08751}, 2022.

\bibitem[Park et~al.(2019)Park, Florence, Straub, Newcombe, and Lovegrove]{park2019deepsdf}
Jeong~Joon Park, Peter Florence, Julian Straub, Richard Newcombe, and Steven Lovegrove.
\newblock Deepsdf: Learning continuous signed distance functions for shape representation.
\newblock In \emph{Proceedings of the IEEE/CVF Conference on Computer Vision and Pattern Recognition (CVPR)}, 2019.

\bibitem[Popovi\'{c} and Hoppe(1997)]{popovic1997progressive}
Jovan Popovi\'{c} and Hugues Hoppe.
\newblock Progressive simplicial complexes.
\newblock In \emph{Proceedings of the 24th Annual Conference on Computer Graphics and Interactive Techniques}, page 217–224, USA, 1997. ACM Press/Addison-Wesley Publishing Co.

\bibitem[Rossignac(1999)]{rossignac2002edgebreaker}
J. Rossignac.
\newblock Edgebreaker: connectivity compression for triangle meshes.
\newblock \emph{IEEE Transactions on Visualization and Computer Graphics}, 5\penalty0 (1):\penalty0 47--61, 1999.

\bibitem[Schwarz et~al.(2022)Schwarz, Sauer, Niemeyer, Liao, and Geiger]{schwarz2022voxgraf}
Katja Schwarz, Axel Sauer, Michael Niemeyer, Yiyi Liao, and Andreas Geiger.
\newblock Voxgraf: Fast 3d-aware image synthesis with sparse voxel grids.
\newblock \emph{Advances in Neural Information Processing Systems}, 35:\penalty0 33999--34011, 2022.

\bibitem[Shen et~al.(2021)Shen, Gao, Yin, Liu, and Fidler]{shen2021deep}
Tianchang Shen, Jun Gao, Kangxue Yin, Ming-Yu Liu, and Sanja Fidler.
\newblock Deep marching tetrahedra: a hybrid representation for high-resolution 3d shape synthesis.
\newblock \emph{Advances in Neural Information Processing Systems}, 34:\penalty0 6087--6101, 2021.

\bibitem[Shim et~al.(2023)Shim, Kang, and Joo]{shim2023diffusion}
Jaehyeok Shim, Changwoo Kang, and Kyungdon Joo.
\newblock Diffusion-based signed distance fields for 3d shape generation.
\newblock In \emph{Proceedings of the IEEE/CVF Conference on Computer Vision and Pattern Recognition (CVPR)}, pages 20887--20897, 2023.

\bibitem[Siddiqui et~al.(2024{\natexlab{a}})Siddiqui, Alliegro, Artemov, Tommasi, Sirigatti, Rosov, Dai, and Nie{\ss}ner]{siddiqui2024meshgpt}
Yawar Siddiqui, Antonio Alliegro, Alexey Artemov, Tatiana Tommasi, Daniele Sirigatti, Vladislav Rosov, Angela Dai, and Matthias Nie{\ss}ner.
\newblock Meshgpt: Generating triangle meshes with decoder-only transformers.
\newblock In \emph{Proceedings of the IEEE/CVF Conference on Computer Vision and Pattern Recognition (CVPR)}, pages 19615--19625, 2024{\natexlab{a}}.

\bibitem[Siddiqui et~al.(2024{\natexlab{b}})Siddiqui, Monnier, Kokkinos, Kariya, Kleiman, Garreau, Gafni, Neverova, Vedaldi, Shapovalov, and Novotny]{siddiqui2024assetgen}
Yawar Siddiqui, Tom Monnier, Filippos Kokkinos, Mahendra Kariya, Yanir Kleiman, Emilien Garreau, Oran Gafni, Natalia Neverova, Andrea Vedaldi, Roman Shapovalov, and David Novotny.
\newblock Meta 3d assetgen: Text-to-mesh generation with high-quality geometry, texture, and pbr materials.
\newblock In \emph{Advances in Neural Information Processing Systems}, pages 9532--9564. Curran Associates, Inc., 2024{\natexlab{b}}.

\bibitem[Takikawa et~al.(2021)Takikawa, Litalien, Yin, Kreis, Loop, Nowrouzezahrai, Jacobson, McGuire, and Fidler]{takikawa2021neural}
Towaki Takikawa, Joey Litalien, Kangxue Yin, Karsten Kreis, Charles Loop, Derek Nowrouzezahrai, Alec Jacobson, Morgan McGuire, and Sanja Fidler.
\newblock Neural geometric level of detail: Real-time rendering with implicit 3d shapes.
\newblock In \emph{Proceedings of the IEEE/CVF Conference on Computer Vision and Pattern Recognition (CVPR)}, pages 11358--11367, 2021.

\bibitem[Tang et~al.(2018)Tang, Dou, Lincoln, Davidson, Guo, Taylor, Fanello, Keskin, Kowdle, Bouaziz, et~al.]{tang2018real}
Danhang Tang, Mingsong Dou, Peter Lincoln, Philip Davidson, Kaiwen Guo, Jonathan Taylor, Sean Fanello, Cem Keskin, Adarsh Kowdle, Sofien Bouaziz, et~al.
\newblock Real-time compression and streaming of 4d performances.
\newblock \emph{ACM Transactions on Graphics (TOG)}, 37\penalty0 (6):\penalty0 1--11, 2018.

\bibitem[Tang et~al.(2020)Tang, Singh, Chou, Hane, Dou, Fanello, Taylor, Davidson, Guleryuz, Zhang, Izadi, Tagliasacchi, Bouaziz, and Keskin]{tang2020deep}
Danhang Tang, Saurabh Singh, Philip~A. Chou, Christian Hane, Mingsong Dou, Sean Fanello, Jonathan Taylor, Philip Davidson, Onur~G. Guleryuz, Yinda Zhang, Shahram Izadi, Andrea Tagliasacchi, Sofien Bouaziz, and Cem Keskin.
\newblock Deep implicit volume compression.
\newblock In \emph{Proceedings of the IEEE/CVF Conference on Computer Vision and Pattern Recognition (CVPR)}, 2020.

\bibitem[Tang et~al.(2024)Tang, Ren, Zhou, Liu, and Zeng]{tang2024dreamgaussian}
Jiaxiang Tang, Jiawei Ren, Hang Zhou, Ziwei Liu, and Gang Zeng.
\newblock Dreamgaussian: Generative gaussian splatting for efficient 3d content creation.
\newblock In \emph{The Twelfth International Conference on Learning Representations}, 2024.

\bibitem[Tang et~al.(2025)Tang, Li, Hao, Liu, Zeng, Liu, and Zhang]{tang2024edgerunner}
Jiaxiang Tang, Zhaoshuo Li, Zekun Hao, Xian Liu, Gang Zeng, Ming-Yu Liu, and Qinsheng Zhang.
\newblock Edgerunner: Auto-regressive auto-encoder for artistic mesh generation.
\newblock In \emph{The Thirteenth International Conference on Learning Representations}, 2025.

\bibitem[Vahdat et~al.(2022)Vahdat, Williams, Gojcic, Litany, Fidler, Kreis, et~al.]{vahdat2022lion}
Arash Vahdat, Francis Williams, Zan Gojcic, Or Litany, Sanja Fidler, Karsten Kreis, et~al.
\newblock Lion: Latent point diffusion models for 3d shape generation.
\newblock \emph{Advances in Neural Information Processing Systems}, 35:\penalty0 10021--10039, 2022.

\bibitem[Wang et~al.(2018)Wang, Zhang, Li, Fu, Liu, and Jiang]{wang2018pixel2mesh}
Nanyang Wang, Yinda Zhang, Zhuwen Li, Yanwei Fu, Wei Liu, and Yu-Gang Jiang.
\newblock Pixel2mesh: Generating 3d mesh models from single rgb images.
\newblock In \emph{Proceedings of the European conference on computer vision (ECCV)}, pages 52--67, 2018.

\bibitem[Wei et~al.(2024)Wei, Zhang, Bi, Tan, Luan, Deschaintre, Sunkavalli, Su, and Xu]{wei2024meshlrm}
Xinyue Wei, Kai Zhang, Sai Bi, Hao Tan, Fujun Luan, Valentin Deschaintre, Kalyan Sunkavalli, Hao Su, and Zexiang Xu.
\newblock Meshlrm: Large reconstruction model for high-quality mesh.
\newblock \emph{arXiv preprint arXiv:2404.12385}, 2024.

\bibitem[Wei et~al.(2025)Wei, Xiang, Bi, Chen, Sunkavalli, Xu, and Su]{wei2023neumanifold}
Xinyue Wei, Fanbo Xiang, Sai Bi, Anpei Chen, Kalyan Sunkavalli, Zexiang Xu, and Hao Su.
\newblock Neumanifold: Neural watertight manifold reconstruction with efficient and high-quality rendering support.
\newblock In \emph{Proceedings of the Winter Conference on Applications of Computer Vision (WACV)}, pages 731--741, 2025.

\bibitem[Weiler(1986)]{weiler1986topological}
Kevin~J Weiler.
\newblock \emph{Topological structures for geometric modeling (Boundary representation, manifold, radial edge structure)}.
\newblock Rensselaer Polytechnic Institute, 1986.

\bibitem[Weng et~al.(2024)Weng, Wang, Zhang, Chen, and Zhu]{weng2024pivotmesh}
Haohan Weng, Yikai Wang, Tong Zhang, CL Chen, and Jun Zhu.
\newblock Pivotmesh: Generic 3d mesh generation via pivot vertices guidance.
\newblock \emph{arXiv preprint arXiv:2405.16890}, 2024.

\bibitem[Weng et~al.(2025)Weng, Zhao, Lei, Yang, Liu, Lai, Chen, Liu, Jiang, Guo, Zhang, Gao, and Chen]{weng2024scaling}
Haohan Weng, Zibo Zhao, Biwen Lei, Xianghui Yang, Jian Liu, Zeqiang Lai, Zhuo Chen, Yuhong Liu, Jie Jiang, Chunchao Guo, Tong Zhang, Shenghua Gao, and C.L.~Philip Chen.
\newblock Scaling mesh generation via compressive tokenization.
\newblock In \emph{Proceedings of the IEEE/CVF Conference on Computer Vision and Pattern Recognition (CVPR)}, pages 11093--11103, 2025.

\bibitem[Wu et~al.(2016)Wu, Zhang, Xue, Freeman, and Tenenbaum]{wu2016learning}
Jiajun Wu, Chengkai Zhang, Tianfan Xue, Bill Freeman, and Josh Tenenbaum.
\newblock Learning a probabilistic latent space of object shapes via 3d generative-adversarial modeling.
\newblock In \emph{Advances in Neural Information Processing Systems}. Curran Associates, Inc., 2016.

\bibitem[Xu et~al.(2024{\natexlab{a}})Xu, Lei, Chen, Zhang, Zhao, Wang, and Tu]{xu2024bayesian}
Haiyang Xu, Yu Lei, Zeyuan Chen, Xiang Zhang, Yue Zhao, Yilin Wang, and Zhuowen Tu.
\newblock Bayesian diffusion models for 3d shape reconstruction.
\newblock In \emph{Proceedings of the IEEE/CVF Conference on Computer Vision and Pattern Recognition (CVPR)}, pages 10628--10638, 2024{\natexlab{a}}.

\bibitem[Xu et~al.(2024{\natexlab{b}})Xu, Cheng, Gao, Wang, Gao, and Shan]{xu2024instantmesh}
Jiale Xu, Weihao Cheng, Yiming Gao, Xintao Wang, Shenghua Gao, and Ying Shan.
\newblock Instantmesh: Efficient 3d mesh generation from a single image with sparse-view large reconstruction models.
\newblock \emph{arXiv preprint arXiv:2404.07191}, 2024{\natexlab{b}}.

\bibitem[Zeng et~al.(2022)Zeng, Vahdat, Williams, Gojcic, Litany, Fidler, and Kreis]{zeng2022lion}
Xiaohui Zeng, Arash Vahdat, Francis Williams, Zan Gojcic, Or Litany, Sanja Fidler, and Karsten Kreis.
\newblock Lion: Latent point diffusion models for 3d shape generation.
\newblock In \emph{Advances in Neural Information Processing Systems}, 2022.

\bibitem[Zhang et~al.(2023)Zhang, Tang, Niessner, and Wonka]{zhang20233dshape2vecset}
Biao Zhang, Jiapeng Tang, Matthias Niessner, and Peter Wonka.
\newblock 3dshape2vecset: A 3d shape representation for neural fields and generative diffusion models.
\newblock \emph{ACM Transactions on Graphics (TOG)}, 42\penalty0 (4):\penalty0 1--16, 2023.

\bibitem[Zhang et~al.(2022)Zhang, Roller, Goyal, Artetxe, Chen, Chen, Dewan, Diab, Li, Lin, et~al.]{zhang2022opt}
Susan Zhang, Stephen Roller, Naman Goyal, Mikel Artetxe, Moya Chen, Shuohui Chen, Christopher Dewan, Mona Diab, Xian Li, Xi~Victoria Lin, et~al.
\newblock Opt: Open pre-trained transformer language models.
\newblock \emph{arXiv preprint arXiv:2205.01068}, 2022.

\bibitem[Zhao et~al.(2025)Zhao, Zhang, Xu, Chen, Xie, Gao, and Tu]{zhao2025depr}
Qingcheng Zhao, Xiang Zhang, Haiyang Xu, Zeyuan Chen, Jianwen Xie, Yuan Gao, and Zhuowen Tu.
\newblock Depr: Depth guided single-view scene reconstruction with instance-level diffusion.
\newblock \emph{arXiv preprint arXiv:2507.22825}, 2025.

\bibitem[Zhao et~al.(2023)Zhao, Liu, Chen, Zeng, Wang, Cheng, FU, Chen, Yu, and Gao]{zhao2023michelangelo}
Zibo Zhao, Wen Liu, Xin Chen, Xianfang Zeng, Rui Wang, Pei Cheng, BIN FU, Tao Chen, Gang Yu, and Shenghua Gao.
\newblock Michelangelo: Conditional 3d shape generation based on shape-image-text aligned latent representation.
\newblock In \emph{Advances in Neural Information Processing Systems}, pages 73969--73982. Curran Associates, Inc., 2023.

\bibitem[Zhou et~al.(2021)Zhou, Du, and Wu]{zhou20213d}
Linqi Zhou, Yilun Du, and Jiajun Wu.
\newblock 3d shape generation and completion through point-voxel diffusion.
\newblock In \emph{Proceedings of the IEEE/CVF International Conference on Computer Vision (ICCV)}, pages 5826--5835, 2021.

\end{thebibliography}
